\documentclass{ringb99}
\usepackage{graphics}

\begin{document}
\title{ X--ray emission from FRII's radio lobes and
 the relativistic particle content}
\author{G. Setti\inst{1,2} 
\and G.Brunetti\inst{1,2} \and A.Comastri\inst{3}}
\institute{Dip. di Astronomia, Univ. di Bologna, via Ranzani 1,
I--40127 Bologna, Italy
\and  
Istituto di Radioastronomia del CNR, via Gobetti 101,
I--40129 Bologna, Italy
\and
Osservatorio Astronomico di Bologna, via Ranzani 1, 
I--40127 Bologna, Italy}

\authorrunning{G.Setti et al.}

\titlerunning{X--rays from FRII's radio lobes}

\maketitle

\begin{abstract}

It is pointed out that copius X-rays from the lobes of FRII radio
galaxies are expected as a natural consequence of the unification linking 
FRIIs and radio-loud quasars. The detection of extended X-ray emission
from 3C 219, likely due to Inverse Compton scattering of the IR photons 
from a hidden quasar with the relativistic electrons in the lobes,
provides the first observational evidence supporting this conjecture.
The X-ray fluxes detected in the directions of other five pwerful FRIIs
with large redshifts may be similarly accounted for (an alternative to
a thermal origin from hot intracluster gas). Much less energetic electrons 
than those producing the synchrotron radio emission are involved in the 
process. Then the equipartition conditions imply
stronger fields, and significantly stronger internal pressures of the
relativic plasmas, than those derived with conventional equipartition
equations. The importance of this effect in the pressure balance with
the confining intergalactic gas is pointed out.

\end{abstract}

\section{Introduction}

At first glance it may appear that the title of this contribution is
somewhat unrelated to the main subject of this workshop. There are, however,
several aspects concerning the physics of strong radio galaxies (FRII) that
may be relevant to the subject being discussed.

First, one can mention the general question of the magnetic field
strengths and of the energy deposited in relativistic 
particles in the lobes of
radio galaxies. This has a direct bearing on the pressure balance with the
ambient gas and, by inference, on the estimates of these quantities
in the radio halos of galaxy clusters. A related aspect is the prediction
of large fluxes of X-rays from the Inverse Compton (IC) process 
in the lobes of extended FRII radio galaxies; this may account for a
sizeable fraction or all of the extended soft X-rays emission
detected in the direction of FRII's at large redshifts and generally
attributed to the presence of surrounding hot intracluster gas.
As a third point it is worth mentioning the possibility of obtaining
important information on the intracluster gas properties from the
interaction with the relativistic plasma in the lobes.
We adopt the unification scheme linking FRII radio galaxies and
radio-loud quasars (Barthel, 1989). Then the radio lobes are flooded
by the radiation from the hidden quasar, leading to very significant
X-ray emission from the IC of the quasar's far/near IR photons with
relativistic electrons having Lorentz factors in the range 100--300
(Brunetti, Setti \& Comastri, 1997; hereafter Paper I). 
The energies of these electrons
are much lower than those typically sampled by the synchrotron radio
emission and, therefore, their presence must be assumed. However,
evidence that this might indeed be the case has been recently gathered
by high angular resolution X-ray observations of 3C 219 with ROSAT-HRI
(Brunetti et al., 1999; hereafter Paper II).

\section{Equipartition Fields}

In general magnetic field intensities in the radio lobes have been 
evaluated assuming the minimum energy condition (equipartition). By
convention this is based on the observed, and extrapoleted as required,
synchrotron spectrum in the 10 MHZ-100 GHz radio band (source frame),
involving electrons with 
$\gamma > 10^3$ in typical fields to be 
found in the lobes. Only
recently the detection of diffuse X-rays from the extended radio lobes 
of Fornax A and Cen B, interpreted as IC scattering of the cosmic
microwave background photons (CMB), has permitted an independent
estimate of the ultra-relativistic electron 
($\gamma > 10^3$) densities 
with the conclusion that the energy in the relativistic particles, the 
energy densities of positive and negative charges being equal, exceeds
that in the magnetic fields (Kaneda et al.1995;  
Tashiro et al.1998). The departure from the equipartition would be further
enhanced with increasing ratio of the energy density in positive charges
(protons) to that in the electron component. Of course, there is no
basic reason why the equipartition argument should strictly apply, except
that it minimizes the energy requirements.

Since the IC production of (soft) X-rays from the scattering of the hidden
quasar IR photons requires relativistic electrons of much lower energies,
it is convenient (and in any case more correct) to work out the minimum
energy condition based on a low energy cut-off ($\gamma_{min}$) in the 
electron energy distribution (Paper I). Ionization losses
may provide a natural mechanism to obtaining a low energy break in the
particle spectra. Under the assumption that the particle energy 
distribution follows a power law with exponent $-\delta$  
and that
particles and fields are uniformely and isotropically distributed,
one finds
that the equipartition field is

\begin{equation}
B_{eq} \simeq 1.3 \cdot 
{B_{eq}^{\prime}}^{ {7\over{\delta+5}} }
\gamma_{min}^{ {{2(2-\delta)}\over{ \delta+5}} }
\end{equation}

where $B^{\prime}_{eq}$ 
is the equipartition field computed with standard 
formulae (Pacholczyk, 1970).
(The ratio between particles and fields energy densities  
depends on the spectral slope, being 1 for 
$\delta = 3$ and the classical 
4/3 for $\delta = 2$).
For $\delta > 2$, 
$B_{eq} > B^{\prime}_{eq}$ 
if $B^{\prime}_{eq} < 1/\gamma_{min}^2$, a condition 
always satisfied in the radio lobes. 
With $\alpha$ in the range 0.8--1.0
it is found that 
$B_{eq}/B^{\prime}_{eq}$ 
is typically comprised in the range 1.5 -- 3.0,
depending on $\gamma_{min}$ and on $B^{\prime}_{eq}$.
Correspondingly 
the normalization of the electron spectrum is 
decreased by at least a factor 2, or larger, an effect
that should be taken into
account in computing the IC emission expected on the basis of the
equipartition value, whereas the ratios 
of the total energy densities
and of the pressures (particles plus fields) is increased by 
a factor $\sim (B_{eq}/B^{\prime}_{eq})^2$. 
The last point is of particular importance not 
only for the evaluation of the minimum energy stored 
in the radio volumes, but
even more so in any discussion concerning the pressure balance with the
surroundig intergalactic gas. 
For instance, by applying the conventional
equipartition relationships it has been found 
(Feretti, Perola \& Fanti, 1992)
that radio tails are underpressurized by factors 
5 -- 10 compared with the
hot confining intracluster gas. 
By considering, as an example, the sources 
with the weakest equipartition fields ($B \sim 1 \mu G$),
and $\gamma_{min} = 20$, we find
that the tail internal pressure should 
be increased by a factor $\sim 8$. 
This may
not fully solve the problem of the pressure balance for the 
large sample of
 sources considered by these authors, 
but it certainly alleviates
the need of strong departures from the equipartition and/or 
the assumption
of large ratios between relativistic nuclei and electrons.

\section{The case of 3C 219}

3C 219 is a well known powerful double-lobed FRII radio galaxy, identified 
with a cD galaxy at a redshift 
z = 0.174 (Clarke et al. 1992 and ref. therein).
The total projected size is $\sim$ 460 Kpc (180 arcsec); 
it has a strong jet
protruding in the direction of the south-western lobe with a projected size
$\sim$ 50 Kpc; its radio spectrum, 
largely dominated by the extended structures,
is a typical power law with spectral index 
$\alpha_r$ = 0.81 in the 178--750 MHz
interval.   
From the combined analysis of ROSAT-PSPC and ASCA archival data
we have concluded (Paper II) that the 0.1-10 keV emission spectrum 
of 3C 219 can be best
represented by a partially ($\sim 74\%$) absorbed power law (photon index 
$\Gamma \simeq  0.75$, absorption column density
$N_H = 2-3 \times 10^{21}$ cm$^{-2}$). 
This could be interpreted as indicating a two
component model: an absorbed nuclear source, with a 
spectrum typical of steep spectrum radio- 
loud quasars, and a scattered component contributing 
$\sim 38\%$ of the total flux
in the 0.1-2.4 keV energy interval 
and with a spectral index $\alpha \simeq 0.8$,
close to the spectral slope measured in the radio band. In addition, from
the spectral analysis we were able to exclude any thermal component 
contributing more than 10\% to the total flux in the ROSAT band.

The above scenario has been largely confirmed by a subsequent deep ROSAT HRI 
(0.1-2.4 keV) observation as described in Paper II. The X-ray image of 3C 219 
is dominated
by a point-like source coincident, within the positional errors, with the
radio core (nucleus) of the galaxy, but a diffuse component extending up
to the hundred kpc scale is also present. Taking into account the results
of the spectral anlysis we find that the point-like source accounts for
$\simeq 60 \%$ 
of the total net counts, while its de-absorbed isotropic luminosity
in the 0.1-2.4 keV band is  
$\simeq 3.6 \times 10^{44}$ 
erg s$^{-1}$, thus strengthening the suggestion
that we are dealing with the emission of a quasar hidden in the nucleus of 
the galaxy in agreement with the unification scheme.

The distribution of the extended emission after subtraction of the point-like
source is shown in Fig.1 superposed on the VLA radio image at 1.4 GHz. The
striking result is the very close coupling between the X-ray and the radio
images. One can distinguish three main components: the central component (C),
which accounts for $>$50\% 
of the extended flux and appears more elongated
in the northern direction along the bottom part of the northern radio lobe,
the south component (S), located mid way between the nucleus and the 
southern hot-spot, and the northern component (N). It is also striking that
the X-ray isophotes appear to carefully avoid the regions of the hot-spots. 

Since we know from the spectral analysis that the X-ray spectrum up to 
10 keV has a slope very close to that measured in tha radio band, it is
tempting to associate the X-ray emission to the IC process involving the
relativistic electrons in the radio lobes. If so, the emission of
component C, whose brightness distribution falls off roughly as the
inverse distance from the nucleus, must be dominated by the IC scattering 
of the IR photons from the hidden quasar with relatively low energy electrons 
($\gamma \sim 100-300$). 
By considering various observational upper limits and the correlations
linking IR, optical, radio and
X-ray radiation properties of quasars and radio galaxies,
we have estimated that the isotropic 6-100 $\mu m$ luminosity
of the hidden quasar could be $5.5 
\times 10^{45}$ erg s$^{-1}$, about an order
of magnitude less than that of a {\it typical} 
radio-loud quasar (see the next
section). We have also included in the photon bath of the electrons the
CMB radiation. We have then constructed an IC model of the source
taking into account the following constraints and assumptions:

$\bullet$
assume a uniform distribution of the relavistic electrons 
throughout the source and the power law energy spectrum of the radio 
electrons ($2\alpha_r + 1$) extrapolated downward to a 
$\gamma_{min}= 50$;

$\bullet$
consider the radiation from the hidden quasar, as seen by the 
electrons,
 made of two components: a direct radiation within two opposite cones of
half-opening angle 45 deg and that re-radiated by the surrounding dusty 
torus;

$\bullet$
assume, according to Bridle et al. (1986), that the radio jet makes an 
angle of 30 deg with respect to the plane of the sky;

$\bullet$
let the orientation of the torus axis be a free parameter subject to the 
constraint that the inner part of component C appears to be elongated
in a direction making a large angle with the radio jet (Fig.1) and that 
the line of sight to the quasar must intercept the torus.
\begin{figure}
 \resizebox{\hsize}{!}{\includegraphics{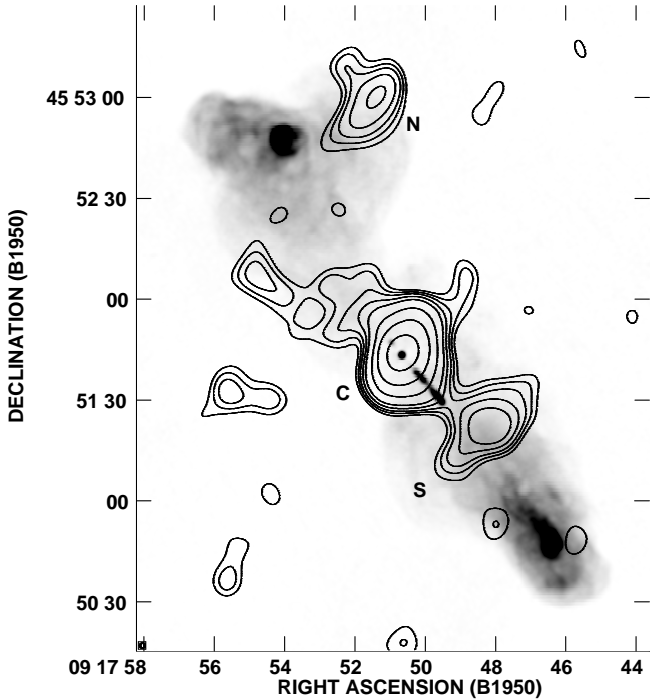}}
\caption[]{The X--ray image (contours) after subtraction
of the central point--like source superposed
on the VLA 1.4 GHz image with $1.4^{\prime\prime}$
resolution (gray--scale).
The contours are : 0.22, 0.24, 0.26, 0.29, 0.33, 0.45,
0.65, 1 cts pixel$^{-1}$ (1 pixel =
2$^{\prime\prime} \times$ 2$^{\prime\prime}$).}
\vskip -0.5cm
\end{figure}
\begin{figure}
 \resizebox{\hsize}{!}{\includegraphics{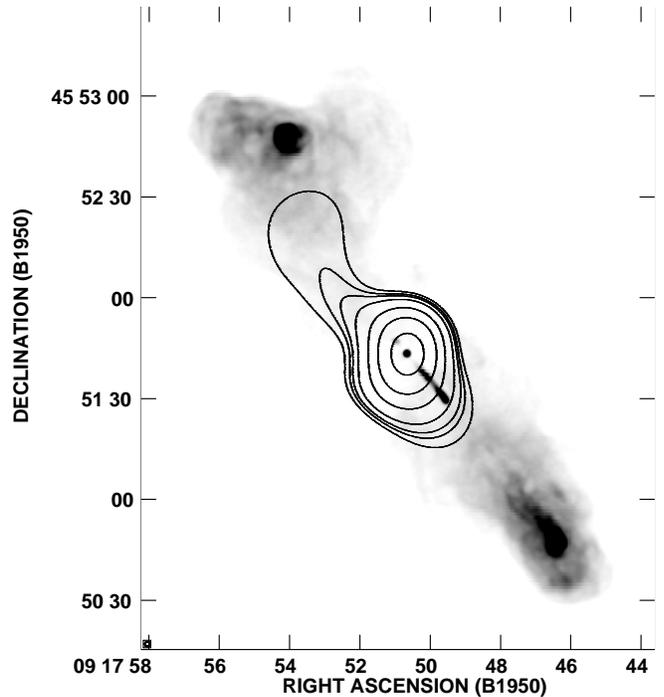}}
\caption[]{The model calculated X--ray isophotes (see text)
are superposed on the VLA 1.4 GHz map (gray--scale).
The isophotes are on the same scale as those of Fig.1, starting
from the 0.24 level.}
\vskip -0.5cm
\end{figure}
The results of our model computations are shown in Fig.2 for the case in 
which the torus axis makes an angle of 
$\sim$ 37 deg with the radio jet and one of
$\sim$73 deg with the line of sight. 
Clearly the model reproduces fairly well
the main features of component C, including the extension of the 
brightness distribution toward the northern lobe due to the dominance
of nearly front scatterings in the far lobe in the observer's direction 
(as discussed in Paper I). The observed flux densities, however, can only
be matched under the assumption that the electron density exceeds the
equipartition value (the particle energy density being equally shared among
the electrons and the nuclei) by about a factor 10, the corresponding magnetic
field strength being 
$\simeq 3 \mu G$. A similarly strong departure from the 
equipartion has been proposed by Tashiro et al.(1998) to explain the X-ray 
emission from Cen B.

We have considered the possibility, also in view of the cluster membership
of 3C 219, that component C could be the result of
a non-symmetric cooling flow. The main point here is that a cooling flow model implies a
thermal contribution far in excess (at least an order of magnitude) of the
upper limit set by the spectral analysis within 2 arcmin radius from the 
source and, in addition, a cluster emission exceeding the ROSAT HRI 
background. We conclude that at this stage our IC model represents the
most solid interpretation of component C, providing at the same time
strong supportive evidence in favour of the unification scheme linking
FRII radio galaxies and radio-loud quasars.

It is worth mentioning that, if the IC interpretation holds, then the
observed coincidence of the X-ray and radio 
spectral slopes implies a 
sufficiently uniform
distribution of particles (with the same spectrum) 
and fields throughout the source. 

Our model is unable to explain the S and N components without introducing
significant changes in the basic assumptions. For instance, both components
can be accounted for by the IC mechanism (mainly from up-scattering
of CMB photons) if one is prepared to admit an 
increase of a factor 2-2.5 in the density of the relativistic electrons at the 
location of the two components. 
An inspection of the radio maps, including
the spectral index distribution and polarization, 
reveals that both components
are located in regions where there is an indication of back-flows from
the hot-spots. It is also possible that there is a thermal contribution 
due to a hot ($kT \simeq 1.5$ keV)
and clumpy intracluster gas surrounding the radio source
in localized regions. 
This is particularly true for the S component whose
almost precise coincidence with a region of moderate depolarization is 
striking. The nature of the emission of these components may soon 
be clarified by upcoming 
observations of 3C 219 with {\it Chandra}. 

\section{
X-ray emission of FRIIs at large redshift}

We briefly report here on a previous work on six strong FRII's (3C 277.2,
280, 294, 324, 356, 368) at large distances 
(z $\simeq$ 0.7 -- 1.8) which had 
been detected by ROSAT PSPC/HRI and discussed in published papers 
(Paper I and ref. therein). These sources show direct or indirect evidence
of extended emission, but appear
very weak typically contributing only several tens of net counts after long
exposure times, so that also their spectra are in general
poorly constrained. The main steps of our procedures were as follows:

$\bullet$ derive a {\it typical} quasar IR-optical spectrum 
(100 $\mu m$ - 350 $nm$) and 
the corresponding (isotropic) luminosity 
$\simeq 9.5 \times 10^{46}$ erg s$^{-1}$ ($M_V \simeq$ -26.1)
typical of high redshift radio-loud quasars;

$\bullet$
for each source assume a relativistic particle power law spectrum from 
the synchrotron radio emission extrapolated downward to a Lorentz factor 
$\gamma_{min}$ = 20 
and derive the equipartion field for equal energy density
of relativistic electrons and protons;

$\bullet$ 
compute the hidden quasar luminosity required to match the observed X-ray
flux of each source by the IC scattering of the quasar photons emitted
within two opposite cones of half opening angle 45 deg coaxial with the
radio axis, including possible corrections due to enhanced anisotropic
IC emission toward the observer for the (unknown) inclination of the radio
axis on the plane of the sky and taking into account the IC X-rays from
the scattering of the CMB photons 
(non negligeable at these large redshifts).
We have concluded that the soft X-ray luminosities and 
spectra of five sources
could be satisfactorely explained by our model, 
with the exception of 3C 324
for which there is independent evidence that the emission is due to a hot 
intracluster gas. Three radio galaxies (3C 277.2, 280, 368) require a hidden
quasar of {\it typical} luminosity; in particular, for 3C 277.2 we 
derived a V magnitude in very good agreement 
with that obtained by Manzini \&
di Serego Alighieri (1996) to explain the optical spectropolarimetric data.
For the remaining two radio galaxies (3C 294, 356) 
the required hidden quasar 
luminosities are larger than {\it typical} (1-1.5 magnitude brighter),
but consistent with 
the indication inferred from spectral and polarimetric data.

\section{Conclusions}

The detection of extended X-ray emission from the FRII radio galaxy 3C 219
can be best interpreted as IC scattering of the IR radiation emitted by
a quasar, hidden in the nucleus of the galaxy, with the relativistic
electrons  residing in the lobes. 
This implies an extension of the electron 
spectrum to energies much lower than those of the synchrotron 
electrons and
a substantial departure from the equipartition, 
that is a factor $\sim 10$
more energy in particles than in the magnetic fields. 
This result provides
further support to the unification scheme linking 
FRIIs and radio-loud quasars.

The soft X-ray fluxes of five strong FRIIs at large redshifts can also
be accounted for as IC scattering of the hidden quasar IR photons with
the relativistic electrons in the radio lobes: under equipartition conditions
the required luminosities of the quasars are very reasonable and consistent 
with spectral and polarization data analysis when available. 
This also
implies that for such distant objects one should be careful 
in ascribing the
observed X-ray emission to a different origin, 
such as a hot intracluster gas
in which these sources might be embedded.

The equipartition values referred to in our analysis have been obtained by
extrapolating the synchrotron electron spectra downward to a minimum cut-off
energy, implying equipartition fields that are significantly stronger than
those derived by the conventional equipartition relationships. 
It follows that
the total energy and internal pressure in the radio 
emitting regions could be 
much larger than usually estimated. 
This should be taken into account while
discussing the pressure balance for the confinement 
of the radio volumes. 
 
\begin{acknowledgements}
This work was partly supported by the Italian Ministry for
University and Research (MURST) under grant Cofin98-02-32.
\end{acknowledgements}

\end{document}